\magnification1200
\baselineskip=1.3\normalbaselineskip

\font\frak=eufm10
\def\g{\hbox{\frak\char'0147}}

\def\s{\hbox{\frak\char'0163}}
\def\p{\hbox{\frak\char'0160}}

\def\l{\hbox{\frak\char'0154}}

\def\h{\hbox{\frak\char'0150}}

\def\1{\hbox{\frak\char'0061}}

\font\block=msbm10
\def\C{\hbox{\block\char'0103}}

\def\R{\hbox{\block\char'0122}}

\def\Z{\hbox{\block\char'0132}}

\def\P{\hbox{\block\char'0120}}
\def\+{\hbox{\block\char'0156}}
\def\H{\hbox{\block\char'0110}}
\def\-{\hbox{\block\char'0157}}

\font\red=eusm10
\def\F{\hbox{\red\char'0106}}

\font\green=msam10
\def\SQ{\hbox{\green\char'0003}}

\font\white=cmbsy10
\def\H1{\hbox{\white\char'0110}}
\def\E1{\hbox{\white\char'0105}}
\def\F1{\hbox{\white\char'0106}}

\font\bigtenrm=cmr10 scaled\magstep1

\font\bf=cmbx10 scaled\magstep0

\font\new=eurm10

\def\p1{\hbox{\new\char'0160}}
\def\h1{\hbox{\new\char'0150}}

\font\frak=eufm10

\font\poisson=eusm10
\def\P1{\hbox{\poisson\char'0120}}

\font\ninerm=cmr9
\noindent
{\bigtenrm On representations of the
exceptional superconformal algebra $CK_6$}

\vskip 0.3in
{\bigtenrm Elena Poletaeva}
\vskip 0.2in

{\it Department of Mathematics}

{\it University of California}

{\it Riverside, CA 92521, USA}

{\it Electronic mail:} elena$@$math.ucr.edu

\vskip 0.4in

{\ninerm   

We realize the exceptional superconformal algebra $CK_6$,
spanned by 32 fields, inside the Lie superalgebra of
pseudodifferential symbols on the supercircle $S^{1|3}$.
We  obtain a one-parameter family of
irreducible representations of $CK_6$ in a superspace spanned by
8 fields.}

\vskip 0.1in
\font\red=eusm10

\noindent{\bf 1. Introduction}
\vskip 0.1in

A {\it superconformal algebra} is a simple complex Lie superalgebra $\g$
spanned by the coefficients of a finite family of
pairwise local fields $a(z) = \sum_{n\in\Z}a_{(n)}z^{-n-1}$,
one of which is the Virasoro field $L(z)$, [3, 8, 11].
Superconformal algebras play an important role in the string theory 
and conformal field theory.

The Lie superalgebras $K(N)$ of contact vector fields with
Laurent polynomials as coefficients (with $N$ odd variables)
is a superconformal algebra which is characterized by
its action on a  contact 1-form, [3, 6, 8, 12].
These Lie superalgebras are also known to physicists 
as the $SO(N)$ superconformal algebras, [1]. 
Note that $K(N)$ is spanned by 
$2^N$ fields. It is simple if $N \not= 4 $, if $N = 4$, then
the derived Lie superalgebra $K'(4)$ is simple.
The nontrivial central extensions of $K(1)$, $K(2)$ and $K'(4)$
are well-known: they are isomorphic to the so-called
Neveu-Schwarz superalgebra, ``the $N = 2$'', and ``the big $N = 4$''
superconformal algebra, respectively, [1].

It was discovered independently in [3] and [17] that
the Lie superalgebra of contact vector fields 
with polynomial coefficients in 1 even and 6 odd variables
contains an exceptional simple Lie superalgebra
(see also [6, 9, 10, 18, 19]).

In [3] a new exceptional superconformal algebra  spanned by 32 fields
was constructed as a subalgebra of $K(6)$, and it was denoted by $CK_6$.
It was proven that $CK_6$ has no nontrivial
central extensions. It was also pointed out that $CK_6$ appears to be
the only new superconformal algebra, which completes their list
(see [11, 12]).

In this work we realize $CK_6$ inside the Poisson superalgebra
of pseudodifferential symbols on the supercircle $S^{1\mid 3}$.
It is known that a Lie algebra of contact vector fields can be realized
as a subalgebra of  Poisson algebra, [2].
In particular, the Lie algebra $Vect(S^1)$ of complex polynomial
vector fields on the circle
has a natural embedding into the Poisson algebra $P$ of formal Laurent series
on the cylinder $T^*S^1\setminus S^1$.
One can consider a family of Lie algebras  $P_{\hbox{h}}$, 
$\hbox{h}\in \rbrack 0, 1]$, having the same underlying vector space,
which contracts to $P$, [13-16].

Analogously, $K(2N)$ is embedded into the Poisson superalgebra 
$P(2N)$ of pseudodifferential symbols on the supercircle 
$S^{1\mid N}$, and there is a family of Lie superalgebras $P_{\hbox{h}}(2N)$,
which contracts to $P(2N)$ (see [20]).

A natural question is whether there exists an embedding 
$$K(2N) \subset P_{\hbox{h}} (2N). \eqno (1. 1)$$
Recall that the answer is ``yes'' if $N = 2$, more precisely,
there exists  an embedding of a nontrivial central extension of
$K'(4) = [K(4), K(4)]$:
$$\hat{K}'(4) \subset  P_{\hbox{h}}(4).\eqno (1. 2)$$
Associated with this embedding, there is a one-parameter family of 
irreducible representations
of $\hat{K}'(4)$ realized on 4 fields, [20].

Note that  embedding (1.1) doesn't hold if $N > 2$, [5].
However, it is remarkable that it is  possible to
embed $CK_6$, which is ``one half'' of $K(6)$,
into $P_{\hbox{h}} (6)$.
In this work we construct this embedding, and obtain the corresponding
one-parameter family of representations 
of $CK_6$ realized on 8 fields.

\vskip 0.1in
\noindent{\bf 2. Contact superconformal algebra $K(2N)$}
\vskip 0.1in

Let $\Lambda(2N)$ be the Grassmann algebra in $2N$ variables
$\xi_1, \ldots, \xi_N, \eta_1, \ldots, \eta_N $, and
let $\Lambda(1, 2N) =\C [t, t^{-1}]\otimes \Lambda (2N)$ be an associative
superalgebra with natural multiplication and with the following parity 
of generators: $p(t) = \bar{0}$, $p(\xi_i) = p(\eta_i) = \bar{1}$
for $i = 1, \ldots, N$. Let
$W(2N)$ be the Lie superalgebra of all derivations of
$\Lambda(1, 2N)$.
Let $\partial_t$, $\partial_{\xi_i}$ and $\partial_{\eta_i}$
stand for $\partial\over {\partial t}$, 
$\partial\over {\partial \xi_i}$ and 
$\partial\over {\partial \eta_i}$, respectively.
By definition,
$$
K(2N) = \lbrace D \in W(2N)\mid D\Omega  = f\Omega \hbox{ for some }
f\in \Lambda(1,2 N)\rbrace, \eqno(2.1)
$$
where
$\Omega = dt + \sum_{i=1}^N \xi_id\eta_i + \eta_id\xi_i$
is a differential 1-form, which is called a {\it contact form}
(see [3, 4, 6, 7, 8, 10, 12]).
The Euler operator is defined by 
$E = \sum_{i=1}^N \xi_i\partial_{\xi_i}
+ \eta_i\partial_{\eta_i}$.
We also define operators $\Delta = 2 - E$ and
$H_f = (-1)^{p(f) +1}\sum_{i=1}^N
\partial_{\xi_i}f
\partial_{\eta_i} + 
\partial_{\eta_i}f
\partial_{\xi_i}$, where $f\in \Lambda(1, 2N)$.

There is a one-to-one correspondence between the differential operators
$D\in K(2N)$ and the functions $f \in \Lambda(1, 2N)$.
The correspondence $f \leftrightarrow D_f$ is given by
$$D_f = \Delta(f){\partial \over {\partial t}} + 
{\partial f \over {\partial t}}E - H_f.\eqno(2.2)$$
The contact bracket on $\Lambda(1, 2N)$ is 
$$\lbrace f, g\rbrace_K = \Delta(f){\partial_t g} -
{\partial_t f}\Delta(g) - \lbrace f, g\rbrace_{P.b},\eqno(2.3)$$
where
$$\lbrace f, g\rbrace_{P.b} = 
(-1)^{p(f) +1}\sum_{i=1}^N \partial_{\xi_i} f
\partial_{\eta_i} g + 
\partial_{\eta_i} f \partial_{\xi_i} g\eqno(2.4)$$
 is the Poisson bracket.
Thus $[D_f, D_g] = D_{{\lbrace f, g \rbrace}_K}$.

The superalgebra $K(6)$ contains an exceptional superconformal algebra,
spanned by 32 fields, as a subalgebra.
This superconformal algebra is denoted by $CK_6$ in [3, 8, 11].
Other notations are also used in the literature (see [6]).
Let $\Theta = \xi_1\xi_2\xi_3\eta_1\eta_2\eta_3$. In what follows 
$(i, j, k) = (1, 2, 3)$ stays for the equality of cyclic permutations.
\hfil\break
{\bf Proposition 1} (see [3, 6]).  $CK_6$ is spanned by the following 32 fields:
$$\eqalignno{
&L_n = t^{n+1} - (\partial_t)^3t^{n+1}\Theta,&(2.5)\cr
&G_n^i = t^{n+1}\xi_i + (\partial_t)^2t^{n+1}\partial_{\eta_i}\Theta,\quad
\tilde{G}_n^i = t^n\eta_i + (\partial_t)^2t^n\partial_{\xi_i}\Theta,
\quad i = 1, 2, 3,\cr
&T_n^{ij} = t^n\xi_i\eta_j - (\partial_t)t^n
\partial_{\eta_i}\partial_{\xi_j}\Theta,\quad i\not= j,\quad
T_n^{i} = t^n\xi_i\eta_i - (\partial_t)t^n
\partial_{\eta_i}\partial_{\xi_i}\Theta, \quad i = 1, 2, 3,\cr
&S_n^{i} = t^n\xi_i(\xi_j\eta_j + \xi_k\eta_k),\quad 
\tilde{S}_n^{i} = t^{n-1}\eta_i(\xi_j\eta_j - \xi_k\eta_k),
\quad i = 1, 2, 3,\cr
&I_n^{i} = t^{n-1}\xi_i\eta_j\eta_k, \quad i = 1, 2, 3,\quad
I_n = t^{n+1}\xi_1\xi_2\xi_3,\cr
&J_n^{ij} = t^{n+1}\xi_i\xi_j - (\partial_t)t^{n+1}
 \partial_{\eta_i}\partial_{\eta_j}\Theta,\quad 
\tilde{J}_n^{ij} = t^{n-1}\eta_i\eta_j - (\partial_t)t^{n-1}
\partial_{\xi_i}\partial_{\xi_j}\Theta,\quad i < j,\cr
}$$
where $n\in \Z$, and $(i, j, k) = (1, 2, 3)$
in the formulae for
$S_n^{i}, \tilde{S}_n^{i}$ and $I_n^{i}$.
\vskip 0.1in
\noindent{\bf 3. The Poisson superalgebra $P(2N)$  of pseudodifferential symbols
on $S^{1\mid N}$}
\vskip 0.1in

The {\it Poisson algebra} $P$ {\it of pseudodifferential symbols
on the circle} is formed by the formal series
$A(t, \xi) = \sum_{-\infty}^na_i(t) {\xi}^i,$
where $a_i(t)\in \C [t, t^{-1}]$, and the variable $\xi$ corresponds
to $\partial_t$. 
The Poisson bracket is defined as follows:
$$
\lbrace A(t, \xi), B(t, \xi) \rbrace = 
\partial_{\xi}A(t, \xi)\partial_{t}B(t, \xi) -
\partial_tA(t, \xi)\partial_{\xi}B(t, \xi). \eqno(3.1)
$$
The Poisson algebra $P$ has a deformation $P_{\hbox{h}}$,
where $\hbox{h}\in \rbrack 0, 1]$.
The associative multiplication in the vector space $P$
is determined as follows:
$$
A(t, \xi)\circ_{\hbox{h}} B(t, \xi) = 
\sum_{n\geq 0} {\hbox{h}^n\over {n!}}\partial^n_{\xi}A(t, \xi)
\partial^n_{t}B(t, \xi).  \eqno(3.2)
$$
The Lie algebra structure on $P_{\hbox{h}}$ is given by
$[A, B]_{\hbox{h}} =  A\circ_{\hbox{h}} B - B\circ_{\hbox{h}} A$, so that the family
$P_{\hbox{h}}$ contracts to $P$. $P_{{\hbox{h}=1}}$ is called 
the  {\it Lie algebra of pseudodifferential symbols
on the circle}, [13-16].

The {\it Poisson superalgebra} $P(2N)$ {\it of 
pseudodifferential symbols on the supercircle}   $S^{1\mid N}$ 
has the underlying vector space
$P\otimes \Lambda (2N)$.
The Poisson bracket is defined as follows:
$$
\lbrace A, B \rbrace = 
\partial_{\xi}A\partial_{t}B -
\partial_tA\partial_{\xi}B +
\lbrace A, B\rbrace_{P.b}.\eqno(3.3)$$

Let $\Lambda_{\hbox{h}}(2N)$ be an associative superalgebra
with generators
$\xi_1, \ldots, \xi_N, \eta_1, \ldots, \eta_N $ and relations:
$\xi_i\xi_j = - \xi_j\xi_i,
\eta_i \eta_j = - \eta_j \eta_i,
\eta_i\xi_j = \hbox{h}\delta_{i,j}  - \xi_j\eta_i.$
Let
$P_{\hbox{h}}(2N) = P_{\hbox{h}} \otimes \Lambda_{\hbox{h}}(2N)$
be an associative superalgebra, where the product of
$A = A_1 \otimes X$ and $B = B_1 \otimes  Y$, 
where $A_1, B_1 \in P_{\hbox{h}}$, and $X, Y \in \Lambda_{\hbox{h}}(2N)$,
is given by
$$AB = {{1}\over \hbox{h}}
(A_1\circ_{\hbox{h}} B_1)\otimes (X Y).\eqno(3.4)$$
Correspondingly, the Lie bracket in $P_{\hbox{h}}(2N)$ is
$[A, B]_{\hbox{h}} = AB - (-1)^{p(A)p(B)}BA$,
and $\hbox{lim}_{{\hbox{h}}\rightarrow 0} [A, B]_{\hbox{h}} = 
\lbrace A, B \rbrace$.
There exist natural embeddings:
$W(N)\subset P(2N)$ and $W(N)\subset P_{\hbox{h}}(2N)$,
where $W(N)$ is the Lie superalgebra of all derivations of
$\C [t, t^{-1}]\otimes \Lambda (\xi_1, \ldots, \xi_N)$,
so that the commutation relations in $P(2N)$ and in $P_{\hbox{h}}(2N)$,
when restricted to $W(N)$, coincide with the commutation relations
in $W(N)$.
$P_{\hbox{h}=1}(2N)$ is called the {\it Lie superalgebra of 
pseudodifferential symbols on}   $S^{1\mid N}$ (see [20]).
\vskip 0.1in
\noindent{\bf 4. Realization of $CK_6$ inside the Poisson superalgebra}
\vskip 0.1in
\noindent
{\bf Theorem 1.} The superalgebra $CK_6$ is spanned
by the following 32 fields inside $P(2N)$:
$$\eqalignno{
&L_{n,0} = t^{n+1}\xi, &(4.1)\cr
&G_{n,0}^i = t^{n+1}\xi\xi_i, \quad 
\tilde{G}_{n,0}^i = t^n\eta_i - nt^{n-1}\xi^{-1}\xi_j\eta_i\eta_j,
\quad i = 1, 2, 3,\cr
&T_{n,0}^{ij} = t^n\xi_i\eta_j - nt^{n-1}\xi^{-1}\xi_k\xi_i\eta_k\eta_j, \quad
i\not= j\not= k,\cr
&T_{n,0}^{i} = -t^n(\xi_j\eta_j + \xi_k\eta_k) +
nt^{n-1}\xi^{-1}\xi_j\xi_k\eta_j\eta_k, \quad i = 1, 2, 3,\cr
&S_{n,0}^{i} = -t^n\xi_i(\xi_j\eta_j + \xi_k\eta_k) + 
nt^{n-1}\xi^{-1}\xi_i\xi_j\xi_k\eta_j\eta_k, \quad i = 1, 2, 3,\cr
&\tilde{S}_{n,0}^{i} = t^{n-1}\xi^{-1}(\xi_j\eta_j - \xi_k\eta_k)\eta_i,
\quad i = 1, 2, 3,\cr
&I_{n,0}^{i} = t^{n-1}\xi^{-1}\xi_i\eta_j\eta_k, \quad i = 1, 2, 3,\quad
I_{n,0} = t^{n+1}\xi\xi_1\xi_2\xi_3,\cr
&J_{n,0}^{ij} = t^{n+1}\xi\xi_i\xi_j, \quad 
\tilde{J}_{n,0}^{ij} = t^{n-1}\xi^{-1}\eta_i\eta_j, \quad i < j,\cr
}$$
where $n\in \Z$, and $(i, j, k) = (1, 2, 3)$
in the formulae for $\tilde{G}_{n,0}^i$, $T_{n,0}^{i}$,
$S_{n,0}^{i}$, $\tilde{S}_{n,0}^{i}$, and $I_{n,0}^{i}$.
\hfil\break
{\it Proof.} Note that there exists an embedding
$$K(2N)\subset P(2N), \quad N\geq 0,\eqno(4.2)$$
see [20]. Consider a $\Z$-grading of the associative superalgebra
$$P(2N) = \oplus_{i\in\Z}P_i(2N)\eqno(4.3)$$
 defined by
$$\eqalignno{
&\hbox{deg } \xi = \hbox{deg } \eta_i = 1, \hbox{ for } i = 1, \ldots, N,&(4.4)\cr
&\hbox{deg } t = \hbox{deg } \xi_i = 0, \hbox{ for } i = 1, \ldots, N.\cr
}$$
With respect to the Poisson bracket,
$$\lbrace P_i(2N), P_j(2N)\rbrace \subset P_{i+j-1}(2N).\eqno(4.5)$$
Thus $P_1(2N)$ is a subalgebra of $P(2N)$, and we will show that
$P_1(2N)\cong K(2N)$.
Equivalently, $P_1(2N)$ is singled out 
as the set of all (Hamiltonian) functions
$A(t, \xi,{\xi}_i, {\eta_i})\in P(2N)$ such that the corresponding vector
fields  supercommute
with the semi-Euler operator:
$$[H_A, \xi\partial_{\xi} 
+ \sum_{i=1}^N{\eta}_i\partial_{\eta_i}] = 0,\eqno(4.6)$$
where
$$A(t, \xi,{\xi}_i, {\eta_i}) \longrightarrow H_A = 
\partial_{\xi} A\partial_t - \partial_t A\partial_{\xi}
-(-1)^{p(A)}\sum_{i=1}^N(\partial_{{\xi}_i}A\partial_{{\eta}_i} +
\partial_{{\eta}_i} A\partial_{{\xi_i}}).\eqno(4.7)$$

To describe an isomorphism from $K(2N)$ onto $P_1(2N)$,
we change the variable $t$ in $\Lambda (1|2N)$:
$t \buildrel\rm \chi\over \rightarrow 2t - \sum_{i = 1}^N\xi_i\eta_i$.
Correspondingly, we have the following contact bracket on $\Lambda (1|2N)$:
$$\lbrace f, g\rbrace_{\tilde{K}} = \tilde{\Delta}(f){\partial_t g} -
{\partial_t f}\tilde{\Delta}(g) - \lbrace f, g\rbrace_{P.b},\eqno(4.8)$$
where
$\tilde{\Delta} = 1 - \tilde{E}$ and
$\tilde{E} = \sum_{i=1}^N \eta_i\partial_{\eta_i}$.
Note that the corresponding contact form is 
$\tilde{\Omega} = dt + \sum_{i=1}^N \xi_id\eta_i$. 
Define a map
$\varphi : \Lambda (1|2N) \rightarrow P_1(2N)$ as follows:
$$f \buildrel\rm\varphi \over 
\rightarrow A_f = (-1)^s\xi^{1-s}f,\eqno(4.9)$$
where $s$ is a scalar given by $\tilde{E}(f) = sf$. Then
$$\lbrace A_f, A_g \rbrace = A_{{\lbrace f, g\rbrace}_{\tilde{K}}}.\eqno(4.10)$$
Applying the isomorphism $\psi = \varphi\circ\chi$ to the fields (2.5), 
we obtain the following fields:
$$\eqalignno{
&\psi(L_n) = 2^{n+1}L_{n,0} - 
2^{n-1}(n+1)(T_{n,0}^{1} + T_{n,0}^{2} + T_{n,0}^{3}), \quad
\psi(G_n^i) = 2^{n+1}G_{n,0}^i - 2^{n}(n+1)S_{n,0}^{i},\cr
&\psi(\tilde{G}_n^i) = -2^{n}\tilde{G}_{n,0}^i + 
2^{n-1}n\tilde{S}_{n,0}^{i},\quad
\psi(T_{n}^{ij}) = -2^{n}T_{n,0}^{ij},\quad
\psi(T_{n}^{i}) = 2^{n-1}(-T_{n,0}^{i} + T_{n,0}^{j} + T_{n,0}^{k}),\cr
&\psi(S_{n}^{i}) = 2^{n}S_{n,0}^{i},\quad
\psi(\tilde{S}_{n}^{i}) = 2^{n-1}\tilde{S}_{n,0}^{i},\quad
\psi(I_{n}^{i}) = 2^{n-1}I_{n,0}^{i},\quad
\psi(I_{n}) = 2^{n+1}I_{n,0},\cr
&\psi(J_{n}^{ij})= 2^{n+1}J_{n,0}^{ij},\quad
\psi(\tilde{J}_{n}^{ij})= 2^{n-1}\tilde{J}_{n,0}^{ij}. &(4.11)\cr
}$$
$$\eqno \SQ$$
\vskip 0.1in
\noindent{\bf 5. Realization of $CK_6$ inside the Lie superalgebra of
pseudodifferential symbols}
\vskip 0.1in
Given the embedding (4.2) it is natural to ask whether there exists
an embedding 
$$K(2N) \subset P_{\hbox{h}} (2N). \eqno (5.1)$$
Recall that if $N = 2$, then there is  an embedding 
$$\hat{K}'(4) \subset  P_{\hbox{h}}(4),\eqno (5.2)$$
where
$K'(4) = [K(4), K(4)]$ is a simple ideal in
$K(4)$ of codimension one defined from the exact sequence
$$0\rightarrow K'(4)\rightarrow K(4)\rightarrow \C 
D_{t^{-1} \xi_1\xi_2\eta_1\eta_2}\rightarrow 0,\eqno (5.3)$$
and $\hat{K}'(4)$ is a nontrivial central extension of $K'(4)$ (see [20]).
The superalgebra $K'(4 )\subset  P(4)$ is spanned by the 12 fields:
$$f(\xi_1, \xi_2, t)\xi  \hbox{ and }    f(\xi_1, \xi_2, t)\eta_i \quad (i = 1, 2),
\eqno (5.4)$$
which form a subalgebra isomorphic to $W(2)$, together with 4 fields: 
$F^i_n$, where $i = 0, 1, 2, 3$, and $n \in \Z$:
$$\eqalignno{
&F^0_n = t^{n-1}\xi^{-1}{\eta}_1{\eta}_2,&(5.5)\cr
&F^i_n = t^{n-1}\xi^{-1}{\xi}_i{\eta}_1{\eta}_2, \quad i = 1, 2, \cr
&F^3_n = t^{n-1}\xi^{-1}{\xi}_1{\xi}_2{\eta}_1{\eta}_2,
\quad n\not= 0.\cr
}$$
\hfil\break
{\bf Proposition 2} ([20]). 
The superalgebra $\hat{K}'(4)$ in (5.2) is spanned by the 
12 fields given in (5.4)
together with 4 fields $F^i_{n,\hbox{h}}$:
$$\eqalignno{
&F^0_{n,\hbox{h}}  = (\xi^{-1}\circ_{\hbox{h}} t^{n-1}){\eta}_1{\eta}_2,&(5.6)\cr
&F^i_{n,\hbox{h}} = (\xi^{-1}\circ_{\hbox{h}} t^{n-1}{\eta}_1{\eta}_2{\xi}_i, \quad i = 1, 2,\cr
&F^3_{n,\hbox{h}} = (\xi^{-1}\circ_{\hbox{h}} t^{n-1}){\eta}_1{\eta}_2{\xi}_1{\xi}_2
+  {\hbox{h}\over{n}}t^{n}, \quad n\not= 0,\cr
}$$
and the central element $\hbox{h}$, so that 
$\hbox{lim}_{\hbox{h}\rightarrow 0}\hat{K}'(4) =  K'(4)\subset P(4)$. 

Note that we cannot obtain the embedding (5.1) if $N > 2$, [5].
However,  the following theorem holds.
\hfil\break
{\bf Theorem 2.} There exists an embedding  $CK_6\subset P_{\hbox{h}}(6)$
for each  $\hbox{h}\in \rbrack 0, 1]$ such that
$\hbox{lim}_{\hbox{h}\rightarrow 0}CK_6 = CK_6 \subset P(6).$
\hfil\break
{\it Proof.} $CK_6$ is spanned by the following fields inside $P_{\hbox{h}}(6)$:
$$\eqalignno{
&L_{n,\hbox{h}} = t^{n+1}\xi, &(5.7)\cr
&G_{n,\hbox{h}}^i = t^{n+1}\xi\xi_i, \quad 
\tilde{G}_{n, \hbox{h}}^i = 
t^n\eta_i - n\xi^{-1}\circ_{\hbox{h}}t^{n-1}\eta_i\eta_j\xi_j,
\quad i = 1, 2, 3, \cr
&T_{n,\hbox{h}}^{ij} = 
t^n\xi_i\eta_j - n\xi^{-1}\circ_{\hbox{h}} t^{n-1}\eta_k\eta_j\xi_k\xi_i,  
\quad i\not= j\not= k,\cr
&T_{n, \hbox{h}}^{i} = -t^n(\xi_j\eta_j + \xi_k\eta_k) +
n\xi^{-1}\circ_{\hbox{h}}t^{n-1}\eta_j\eta_k\xi_j\xi_k + \hbox{h}t^n,
\quad i = 1, 2, 3,\cr
&S_{n, \hbox{h}}^{i} = -t^n\xi_i(\xi_j\eta_j + \xi_k\eta_k) + 
n\xi^{-1}\circ_{\hbox{h}} t^{n-1}\eta_j\eta_k\xi_i\xi_j\xi_k + 
\hbox{h}t^n\xi_i, \quad i = 1, 2, 3,\cr
&\tilde{S}_{n, \hbox{h}}^{i} = \xi^{-1}\circ_{\hbox{h}}t^{n-1}
(\eta_j\eta_i\xi_j - \eta_k\eta_i\xi_k), \quad
i = 1, 2, 3,\cr
&I_{n, \hbox{h}}^{i} = \xi^{-1}\circ_{\hbox{h}} t^{n-1}\eta_j\eta_k\xi_i, \quad
i = 1, 2, 3,\quad
I_{n, \hbox{h}} = t^{n+1}\xi\xi_1\xi_2\xi_3,\cr
&J_{n, \hbox{h}}^{ij} = t^{n+1}\xi\xi_i\xi_j, \quad 
\tilde{J}_{n, \hbox{h}}^{ij} = \xi^{-1}\circ_{\hbox{h}}t^{n-1}\eta_i\eta_j,
\quad i < j,\cr
}$$
where $n\in\Z$, and $(i, j, k) = (1, 2, 3)$ 
in the formulae for 
$\tilde{G}_{n, \hbox{h}}^i$, $T_{n, \hbox{h}}^{i}$,
$S_{n, \hbox{h}}^{i}$, $\tilde{S}_{n, \hbox{h}}^{i}$ and $I_{n, \hbox{h}}^i$.
Let $\hbox{h}\in [0, 1]$. Set
$J_{n, \hbox{h}}^{ij} = - J_{n, \hbox{h}}^{ji}$ and
$\tilde{J}_{n, \hbox{h}}^{ij} = - \tilde{J}_{n, \hbox{h}}^{ji}$ for $i > j$.
Given $\hbox{h}\in [0, 1]$, set
$$L_n := L_{n,\hbox{h}}, \quad\ldots, \quad 
\tilde{J}_{n}^{ij} := \tilde{J}_{n, \hbox{h}}^{ij}.\eqno (5.8)$$
Recall that if $\hbox{h} = 0$, then (5.8) gives elements (4.1).
The nonvanishing commutation relations 
between the elements   (5.8) 
are as follows:
let $i\not= j\not= k$, then
$$\eqalignno{
&[L_n, L_m] = (m - n)L_{n+m},
[L_n, G_m^i] = (m - n)G_{n+m}^i,
[L_n, \tilde{G}_m^i] = m\tilde{G}_{n+m}^i, &(5.9)\cr
&[L_n, T_m^{ij}] = mT_{n+m}^{ij},
[L_n, T_m^{i}] = mT_{n+m}^{i},
[L_n, S_m^{i}] = mS_{n+m}^{i},
[L_n,\tilde{S}_m^{i}] = (m+n)\tilde{S}_{n+m}^{i},\cr
&[L_n, I_m^{i}] = (m + n)I_{n+m}^{i},
[L_n, I_m] = (m - n)I_{n+m},
[L_n, J_m^{ij}] = (m-n)J_{n+m}^{ij},\cr
&[L_n, \tilde{J}_m^{ij}] = (m+n)\tilde{J}_{n+m}^{ij},
[G_n^i, G_m^j] = (m-n)J_{n+m}^{ij},
[G_n^i,\tilde{G}_m^j] = mT_{n+m}^{ij},\cr
&[G_n^i, T_m^{ji}] = - G_{n+m}^j + mS_{n+m}^{j},
[G_n^i, T_m^{i}] = mS_{n+m}^{i},
[G_n^i, T_m^{j}] = G_{n+m}^i,\cr
&[G_n^i, S_{m}^j] =  J_{n+m}^{ij},
[G_n^i, \tilde{S}_{m}^j] =  T_{n+m}^{ij},
[\tilde{G}_n^i, \tilde{G}_m^j] = (m-n)\tilde{J}_{n+m}^{ij},\cr
&[\tilde{G}_n^i, S_{m}^i] =  T_{n+m}^{i},
[\tilde{G}_n^i, S_{m}^j] = T_{n+m}^{ji},
[\tilde{G}_n^i, \tilde{S}_{m}^j] = -\tilde{J}_{n+m}^{ij},
[\tilde{G}_n^i, J_{m}^{ij}] = G_{n+m}^j,\cr
&[T_n^{ij}, T_m^{ji}] = T_{n+m}^{i} - T_{n+m}^{j},
[T_n^{ij}, T_m^{jk}] =  T_{n+m}^{ik},
[T_n^{ij}, T_m^{ki}] =  -T_{n+m}^{kj},
[T_n^{ij}, T_m^{i}] = -T_{n+m}^{ij},\cr
&[T_n^{ij}, T_m^{j}] = T_{n+m}^{ij},
[T_n^{ij}, S_m^{j}] = S_{n+m}^{i},
[T_n^{ij}, \tilde{S}_m^{i}] = \tilde{S}_{n+m}^{j},
[T_n^{ij}, \tilde{S}_m^{k}] = -2I_{n+m}^{i},\cr
&[T_n^{ij}, I_m^{j}] = -\tilde{S}_{n+m}^{k},
[T_n^{ij}, J_m^{jk}] = J_{n+m}^{ik},
[T_n^{ij}, \tilde{J}_m^{ik}] = -\tilde{J}_{n+m}^{jk},
[T_n^j, S_m^i] = -S_{n+m}^i,\cr
&[T_n^j, \tilde{S}_m^i] = \tilde{S}_{n+m}^i,
 [T_n^i, I_m^i] = 2I_{n+m}^i, 
[T_n^i, I_m] = -2I_{n+m},\cr
&[T_n^i, J_{m}^{ij}] = [T_n^j, J_{m}^{ij}] = - J_{n+m}^{ij},
[T_n^k, J_{m}^{ij}] = -2 J_{n+m}^{ij},
[T_n^i, \tilde{J}_{m}^{ij}] = [T_n^j, \tilde{J}_{m}^{ij}]
 = \tilde{J}_{n+m}^{ij},\cr 
&[T_n^k, \tilde{J}_{m}^{ij}] = 2\tilde{J}_{n+m}^{ij},
[J_{n}^{ij}, \tilde{J}_{m}^{ij}] = T_{n+m}^k,
[J_{n}^{ij}, \tilde{J}_{m}^{ik}] = -T_{n+m}^{jk},
[J_{n}^{ij}, \tilde{J}_{m}^{jk}] =  T_{n+m}^{ik}.\cr
}$$
Let $(i, j, k) = (1, 2, 3)$, then
$$\eqalignno{
&[G_n^i, \tilde{G}_m^i] = L_{n+m} -mT_{n+m}^{k},
[G_n^i, \tilde{S}_{m}^i] = T_{n+m}^{j} - T_{n+m}^{k},
[G_n^i, I_{m}^j] = T_{n+m}^{jk}, 
[G_n^i, I_{m}^k] = -T_{n+m}^{kj},\cr
&[G_n^i, J_{m}^{jk}] = (m-n)I_{n+m},
[G_n^i, \tilde{J}_{m}^{jk}] = (m+n)I_{n+m}^i,
[G_n^i, \tilde{J}_{m}^{ij}] = \tilde{G}_{n+m}^j - (n+m)\tilde{S}_{n+m}^{j},\cr
&[G_n^i, \tilde{J}_{m}^{ik}] =  \tilde{G}_{n+m}^{k},
[\tilde{G}_n^i, T_m^{ij}] =  \tilde{G}_{n+m}^j -n\tilde{S}_{n+m}^{j},
[\tilde{G}_n^i, T_m^{ik}] = \tilde{G}_{n+m}^k - (n+m)\tilde{S}_{n+m}^{k},\cr
&[\tilde{G}_n^i, T_m^{jk}] = (m+n)I^j_{n+m},
[\tilde{G}_n^i, T_m^{kj}] = (n-m)I^k_{n+m},
[\tilde{G}_n^i, T_m^{j}] = -\tilde{G}_{n+m}^i +m\tilde{S}_{n+m}^i,\cr
&[\tilde{G}_n^i, T_m^{k}] = -\tilde{G}_{n+m}^i,
[\tilde{G}_n^i, I_{m}^i] = \tilde{J}_{n+m}^{jk},
[\tilde{G}_n^i, I_{m}] = J_{n+m}^{jk},
[S_n^i, {J}_{m}^{jk}] = -2I_{n+m},
[S_n^i, \tilde{J}_{m}^{ij}] = -\tilde{S}_{n+m}^{j},\cr
&[S_n^i, \tilde{J}_{m}^{ik}] = \tilde{S}_{n+m}^{k},
[S_n^i, \tilde{J}_{m}^{jk}] = 2I_{n+m}^{i},
[\tilde{S}_n^i, J_{m}^{ij}] = {S}_{n+m}^j,
[\tilde{S}_n^i, J_{m}^{ik}] = -{S}_{n+m}^k,
[I_n^i, J_{m}^{jk}] = -{S}_{n+m}^i,\cr
&[I_n, \tilde{J}_{m}^{ij}] = {S}_{n+m}^k. &(5.10)\cr
}$$
$$\eqno \SQ$$
\vskip 0.1in
\noindent{\bf 6. Representation of $CK_6$ associated with its embedding
into $P_{\hbox{h=1}}(6)$}
\vskip 0.1in
Recall that the embedding (5.2) for $\hbox{h} = 1$ allows to define
a one-parameter family of
spinor-like representations of $K'(4)$ in the superspace spanned by 2 even and 2
odd fields, where the central element $\hbox{h}$ acts by the identity operator,
[20].
\hfil\break
{\bf Theorem 3.} There exists a one-parameter family of irreducible 
representations of $CK_6$, depending on parameter $\mu\in \C$,
in a superspace spanned by 4 even fields and 4 odd fields.
\hfil\break
{\it Proof.}
Let $V^{\mu} =  t^{\mu} \C [t, t^{-1}]\otimes \Lambda (3)$, where
$\Lambda (3) = \Lambda (\xi_1, \xi_2, \xi_3)$ is the Grassmann algebra,
and $\mu \in \R \setminus\Z$.
Let $\lbrace v_m^i, \hat{v}_m^i\rbrace$, where ${m\in \Z}$ and 
$i =  1, 2, 3, 4$,
be the following basis in $V^{\mu}$:
$$v_m^i = {t^{m+\mu}\over {m+\mu}}\xi_i,\quad
\hat{v}_m^i = t^{m+\mu}\xi_j\xi_k,\quad i = 1, 2, 3,\quad
v_m^4 = {t^{m+\mu}\over {m+\mu}},\quad\hat{v}_m^4 = -t^{m+\mu}\xi_1\xi_2\xi_3,
\eqno (6.1)$$
where $(i, j, k) = (1, 2, 3)$ in the formulae for $\hat{v}_m^i$.
We define a representation of $CK_6$ in $V^{\mu}$
according to the formulae (5.7), where $\hbox{h} =1$.
Namely, $\xi_i$ is the operator of multiplication in
$\Lambda (3)$, $\eta_i$ is identified with $\partial_{\xi_i}$, and 
$\xi^{-1}$ is identified with the anti-derivative:
$$\xi^{-1}g(t) = \int g(t)dt, \quad g\in   t^{\mu} \C [t, t^{-1}].
\quad \eqno (6.2)$$ 
Notice that  the  formula
$$\xi^{-1}\circ_{\hbox{h=1}} f= \sum_{n=0}^{\infty}(-1)^n(\xi^{n}f)\xi^{-n-1},
\eqno (6.3)$$
where $f\in \C [t, t^{-1}]$, when applied to a function 
$g\in t^{\mu} \C [t, t^{-1}]$,
corresponds to the formula of integration by parts:
$$\int fgdt = f \int  gdt - f'\int\int gdt^2 + f''\int\int\int gdt^3 
-\ldots. \eqno (6.4)$$
The superalgebra $CK_6$ acts on $V^{\mu}$ as follows (see (5.8) for notations):
$$\eqalignno{
&L_{n}(v_m^i) = (m + n + \mu)v_{m+n}^i, \quad
L_{n}(\hat{v}_m^i) = (m + \mu)\hat{v}_{m+n}^i, &(6.5)\cr
&G_{n}^i (v_m^4) = (m + n + \mu)v_{m+n}^i,\quad
G_{n}^i(\hat{v}_m^i) = -(m +\mu)\hat{v}_{m+n}^4,\cr
&G_{n}^i (v_m^{j}) = \hat{v}_{m+n}^k,\quad
G_{n}^i (v_m^{k}) = -\hat{v}_{m+n}^j,\quad
\tilde{G}_{n}^i(v_m^{i}) = v_{m+n}^{4}, \quad
\tilde{G}_{n}^i(\hat{v}_m^4) = -\hat{v}_{m+n}^i,\cr
&\tilde{G}_{n}^i(\hat{v}_m^j) = -(m + \mu){v}_{m+n}^k,\quad
\tilde{G}_{n}^i(\hat{v}_m^k) = (m + n  + \mu){v}_{m+n}^j,\cr
&T_{n}^{ij}(v_m^{j}) = v_{m+n}^{i},\quad
T_{n}^{ij}(\hat{v}_m^i) = -\hat{v}_{m+n}^j,\quad
T_{n}^{i}(v_m^{i}) =  v_{m+n}^{i},\quad
T_{n}^{i}(v_m^4) = v_{m+n}^4,\cr
&T_{n}^{i}(\hat{v}_m^i) = - \hat{v}_{m+n}^i\quad
T_{n}^{i}(\hat{v}_m^4) = - \hat{v}_{m+n}^4,\quad
S_{n}^{i}(v_m^{4}) = v_{m+n}^{i}, \quad
S_{n}^{i}(\hat{v}_m^i) = \hat{v}_{m+n}^4,\cr
&\tilde{S}_{n}^{i} (\hat{v}_m^j) = v_{m+n}^{k},\quad
\tilde{S}_{n}^{i} (\hat{v}_m^k) = v_{m+n}^{j}, \quad
I_{n}^{i}(\hat{v}_m^i) = -v_{m+n}^{i},\quad
I_{n}(v_m^{4}) = -\hat{v}_{m+n}^4,\cr
&J_{n}^{ij}(v_m^{4}) = \hat{v}_{m+n}^k,\quad
J_{n}^{ij}(v_m^{k}) = -\hat{v}_{m+n}^4,\quad
\tilde{J}_{n}^{ij}(\hat{v}_m^k) = -v_{m+n}^{4},\quad
\tilde{J}_{n}^{ij}(\hat{v}_m^4) = v_{m+n}^{k},\cr
}$$
where $(i, j, k) = (1, 2, 3)$ in the formulae for
$\tilde{G}_{n}^i$, $\tilde{S}_{n}^{i}$, $J_{n}^{ij}$, and
$\tilde{J}_{n}^{ij}$.
Formulae (6.5) define a one-parameter family of representations of 
$CK_6$ in $V^{\mu} = < v_m^i, \hat{v}_m^i \hbox{ } |\hbox{ }
i = 1, \ldots, 4, \hbox{ } m\in\Z>$.
$$\eqno \SQ$$
{\bf Remark 1.} We have posed the condition 
$\mu \in \R \setminus\Z$ in the definition of $V^{\mu}$.
However, formulae (6.5) actually define a representation of $CK_6$
in a superspace spanned by $v_m^i, \hat{v}_m^i$ for an arbitrary $\mu\in \C$.
(See also section 8).

\vskip 0.1in
\noindent{\bf 7. The second family of representations of $CK_6$}
\vskip 0.1in

Note that the embedding of infinite-dimensional Lie superalgebras
$$CK_6 \subset K(6),\eqno (7.1)$$
considered in this work, is naturally related to the embedding
of finite-dimensional Lie superalgebras
$$\hat{\P1}(4) \subset P(0|6).\eqno (7.2)$$
Recall that $P(0|6)$ is the Poisson superalgebra with 6 odd generators:
$\xi_1,\xi_2, \xi_3, \eta_1, \eta_2, \eta_3$, and the Poisson bracket
is given by (2.4).
The simple Lie superalgebra $\P1(n)$ is defined as follows.
Let $\tilde{\P1}(n)$ be the Lie superalgebra, which preserves 
the odd nondegenerate 
supersymmetric bilinear form antidiag $(1_n, 1_n)$
on the $(n|n)$-dimensional superspace. Thus
$$\tilde{\P1}(n) = \lbrace \pmatrix{A&B\cr
C&-A^t\cr}\hbox{ }|\hbox{ }A\in \g\l(n),
B \hbox{ and } C \hbox{ are }
n\times n -\hbox{ matrices}, B^t = B, C^t = -C\rbrace.\eqno (7.3)$$
${\P1}(n)$ is a subalgebra of $\tilde{\P1}(n)$ such that 
$A\in \s\l(n)$, [7]. A. Sergeev has proved that
${\P1}(n)$ has a nontrivial central extension if and only if $N = 4$,
see [18]. Note that $\hbox{dim }\hat{\P1}(4) = (16|16)$.
It was pointed out in [6, 18] that $\hat{\P1}(4)$ has a family 
$\hbox{spin}_{\lambda}$ of
$(4|4)$-dimensional irreducible representations.
In fact, there exist two such families:
they correspond to two families
of embeddings of $\hat{\P1}(4)$ into $P(0|6)$.

For every $\lambda\not= 0$ we can realize $\hat{\P1}(4)$
inside $P(0|6)$ as follows:

$$\hat{\P1}(4) = <L, G^i, \tilde{G}^i, T^{ij}, T^{i},
S^{i}, \tilde{S}^{i}, I^{i}, I, J^{ij}, \tilde{J}^{ij}>,
\eqno (7.4)$$ 
where
$$\eqalignno{
&L = \lambda, G^i = \lambda\eta_i, \tilde{G}^i = \xi_i,
T^{ij} = \eta_i\xi_j, T^{i} = -\eta_j\xi_j - \eta_k\xi_k, &(7.5)\cr
&S^{i} = -\eta_i(\eta_j\xi_j + \eta_k\xi_k),
\tilde{S}^{i} = {1\over{\lambda}} (\eta_j\xi_j - \eta_k\xi_k)\xi_i,\cr
&I^{i} = {1\over{\lambda}} \eta_i\xi_j\xi_k, I = \lambda \eta_1\eta_2\eta_3,
J^{ij} = \lambda\eta_i\eta_j, \tilde{J}^{ij} = {1\over{\lambda}}\xi_i\xi_j,\cr
}$$
so that the central element is $L$.
Correspondingly, there is an embedding of $\hat{\P1}(4)$ into $P_{\hbox{h}}(0|6)$
given by
$$\eqalignno{
&L_{\hbox{h}} = \lambda, G_{\hbox{h}}^i = \lambda\eta_i, 
\tilde{G}_{\hbox{h}}^i = \xi_i,
T_{\hbox{h}}^{ij} = \eta_i\xi_j, 
T_{\hbox{h}}^{i} = -\eta_j\xi_j - \eta_k\xi_k + {\hbox{h}}, &(7.6)\cr
&S_{\hbox{h}}^{i} = -\eta_i(\eta_j\xi_j + \eta_k\xi_k) + {\hbox{h}}\eta_i,
\tilde{S}_{\hbox{h}}^{i} = {1\over{\lambda}}
(\xi_j\xi_i\eta_j - \xi_k\xi_i\eta_k),\cr
&I_{\hbox{h}}^{i} = {1\over{\lambda}} \xi_j\xi_k\eta_i, I_{\hbox{h}} = \lambda \eta_1\eta_2\eta_3,
J_{\hbox{h}}^{ij} = \lambda\eta_i\eta_j, \tilde{J}_{\hbox{h}}^{ij} = 
{1\over{\lambda}}\xi_i\xi_j,\cr
}$$
and $\lim_{\hbox{h}\rightarrow 0}\hat{\P1}(4) = \hat{\P1}(4)\subset P(0|6)$.
The nonvanishing commutation relations 
between the elements (7.5) and 
between the elements (7.6) are as in (5.9)-(5.10), where the indexes
  $m = n = 0$.

Associated to this embedding (for $\hbox{h} = 1$) there is a family 
$\hbox{spin}_{\lambda}^1$ of representations of $\hat{\P1}(4)$
in the superspace $\Lambda (\xi_1, \xi_2, \xi_3)$. We choose the basis 
$$v^i = \xi_i, \hbox{ }\hat{v}^i = {1\over {\lambda}}\xi_j\xi_k, \hbox{ }i = 1,2 ,3,
\hbox{ } v^4 = 1, \hbox{ }\hat{v}^4 = -{1\over {\lambda}} \xi_1\xi_2\xi_3.
\eqno (7.7)$$ Explicitely,
$$\hbox{spin}_{\lambda}^1:
\pmatrix{A&B\cr
C&-A^t\cr} + \C L \rightarrow \pmatrix{A&B-\lambda\tilde{C}\cr
C&-A^t\cr} + \C\lambda\cdot1_{4|4},\eqno (7.8)$$
where $1_{4|4}$ is the identity matrix,
and if $C_{ij} = E_{ij} - E_{ji}$, then $\tilde{C}_{ij} = {C}_{kl}$,
so that the permutation $(1, 2, 3, 4)\mapsto (i, j, k, l)$ is even,
cf. [6, 18]. Formula (7.8) also gives the standard representation
$\hbox{spin}_{0}^1$.

The second family of embeddings of $\hat{\P1}(4)$ into $P(0|6)$
and into  $P_{\hbox{h}}(0|6)$
is given by (7.4)-(7.6), where  $\xi_i$ is interchanged with $\eta_i$
for all $i$ in all the formulae. Correspondingly,
there is a family 
$\hbox{spin}_{\lambda}^2$ of representations of $\hat{\P1}(4)$
associated to this embedding (for $\hbox{h} = 1$) 
in the superspace $\Lambda (\xi_1, \xi_2, \xi_3)$, so that
$\Pi(\hbox{spin}_{\lambda}^2) \cong \hbox{spin}_{\lambda}^1$,
as $\hat{\P1}(4)$-modules,
for all $\lambda$.
\hfil\break
($\Pi$ denotes the change of parity).
From Theorem 3 we have the following corollary.
\hfil\break
{\bf Corollary 1.}
Under the restriction of the representation of
$CK_6$ in $V^{\mu}$ to $\hat{\P1}(4)$,
$V^{\mu}$ decomposes into a direct sum of irreducible 
$(4|4)$-dimensional representations of the family $\hbox{spin}_{\lambda}^2$.
\hfil\break
{\it Proof.} 
Naturally, there are embeddings:
$$\hat{\P1}(4) \subset CK_6, \quad P(0|6)\subset K(6).\eqno (7.9)$$
The first embedding is given as follows:
$$\hat{\P1}(4) = \lbrace x\in CK_6 \hbox{ }|\hbox{ } [L_0, x] = 0\rbrace,
\eqno (7.10)$$
hence $L_0$ is the central element.
The nontrivial 2-cocycle on ${\P1}(4)$ is 
$(G_{0}^i, \tilde{G}_{0}^j) = \delta_{i, j}L_0$.
It follows from (6.5) that $V^{\mu}$ is a direct sum of 
$(4|4)$-dimensional $\hat{\P1}(4)$-submodules:
$$V^{\mu} = \oplus_{m\in \Z}V^{\mu}_m, \quad 
V^{\mu}_m = <v_m^i, \hat{v}_{m}^i \hbox{ } | \hbox{ } i = 1, 2, 3, 4 >,
\eqno (7.11)$$
where $V^{\mu}_m \cong \hbox{spin}_{m + \mu}^2$.
$$\eqno \SQ$$

It is possible to define another embedding of
$CK_6$ into $P(6)$ (respectively, into $P_{\hbox{h}}(6)$)
by interchanging $\xi_i$ with $\eta_i$ in all the formulae
(4.1) (respectively in (5.7)), 
and then obtain a one-parameter family
of representations of $CK_6$  in $V^{\mu}$
by repeating the previous construction. Thus the following theorem holds.
\hfil\break
{\bf Theorem 4.} Consider the following basis in $V^{\mu}$:
$$v_m^i = t^{m+\mu}\xi_i,\quad
\hat{v}_m^i = {t^{m+\mu}\over{m+\mu}} \xi_j\xi_k,\quad i = 1, 2, 3,\quad
v_m^4 = t^{m+\mu},\quad \hat{v}_m^4 = -{t^{m+\mu}\over{m+\mu}}\xi_1\xi_2\xi_3,
\eqno (7.12)$$
where $(i, j, k) = (1, 2, 3)$ in the formulae for $\hat{v}_m^i$.
Then the action of $CK_6$ on 
$V^{\mu}$
is defined as follows
$$\eqalignno{
&L_{n}(v_m^i) = (m + \mu)v_{m+n}^i, \quad
L_{n}(\hat{v}_m^i) = (m + n + \mu)\hat{v}_{m+n}^i, &(7.13)\cr
&G_{n}^i (v_m^i) = (m + \mu)v_{m+n}^4,\quad
G_{n}^i(\hat{v}_m^4) = -(m + n + \mu)\hat{v}_{m+n}^i,\cr
&G_{n}^i (\hat{v}_m^{k}) = {v}_{m+n}^j,\quad
G_{n}^i (\hat{v}_m^{j}) = -{v}_{m+n}^k,\quad
\tilde{G}_{n}^i(v_m^{4}) = {v}_{m+n}^i,\quad
\tilde{G}_{n}^i(\hat{v}_m^{i}) = -\hat{v}_{m+n}^4,\quad\cr
&\tilde{G}_{n}^i(v_m^{k}) = -(m + n  + \mu)\hat{v}_{m+n}^{j}, \quad
\tilde{G}_{n}^i({v}_m^j) =  (m + \mu) \hat{v}_{m+n}^k,\cr
&T_{n}^{ij}(v_m^{i}) = -v_{m+n}^{j},\quad
T_{n}^{ij}(\hat{v}_m^j) = \hat{v}_{m+n}^i,\quad
T_{n}^{i}(v_m^{i}) =  -v_{m+n}^{i},\quad
T_{n}^{i}(v_m^4) = -v_{m+n}^4,\cr
&T_{n}^{i}(\hat{v}_m^i) =  \hat{v}_{m+n}^i\quad
T_{n}^{i}(\hat{v}_m^4) =  \hat{v}_{m+n}^4,\quad
S_{n}^{i}(v_m^{i}) = -v_{m+n}^{4}, \quad
S_{n}^{i}(\hat{v}_m^4) = -\hat{v}_{m+n}^i,\cr
&\tilde{S}_{n}^{i} ({v}_m^k) = -\hat{v}_{m+n}^{j},\quad
\tilde{S}_{n}^{i} ({v}_m^j) = -\hat{v}_{m+n}^{k}, \quad
I_{n}^{i}({v}_m^i) = \hat{v}_{m+n}^{i},\quad
I_{n}(\hat{v}_m^{4}) = {v}_{m+n}^4,\cr
&J_{n}^{ij}(\hat{v}_m^{k}) = -{v}_{m+n}^4,\quad
J_{n}^{ij}(\hat{v}_m^{4}) = {v}_{m+n}^k,\quad
\tilde{J}_{n}^{ij}({v}_m^4) = \hat{v}_{m+n}^{k},\quad
\tilde{J}_{n}^{ij}({v}_m^k) = -\hat{v}_{m+n}^{4},\cr
}$$
where $(i, j, k) = (1, 2, 3)$ in the formulae for
$\tilde{G}_{n}^i$, $\tilde{S}_{n}^{i}$, $J_{n}^{ij}$, and
$\tilde{J}_{n}^{ij}$.
Thus
$V^{\mu}$ is a direct sum of 
$(4|4)$-dimensional $\hat{\P1}(4)$-submodules, see (7.11),
where $V^{\mu}_m \cong \hbox{spin}_{m+\mu}^1$.
\vskip 0.1in
\noindent
{\bf 8. Final remarks}

In Theorem 1 we realized  $CK_6$ inside the deg = 1 part of
the $\Z$-grading of $P(6)$, given by (4.3), and 
in Theorem 2 we realized  $CK_6$ inside $P_{\hbox{h}}(6)$.
One should note that in this realization the elements of $CK_6$ have powers 
$-1, 0$ and  $1$
with respect to $\xi$, see (4.1) and (5.7).

We will now show how to single out $CK_6$ from $P_{\hbox{h}}(6)$.
Let $S$ be a subspace of $P_{\hbox{h}}(6)$ spanned by
$W(3)$ (which consists of the elements of power $0$ and  $1$
with respect to $\xi$) 
and the following fields $(n \in \Z$):
$$\eqalignno{
&\xi^{-1}\circ_{\hbox{h}}t^{n-1}\eta_i\eta_j,\quad
\xi^{-1}\circ_{\hbox{h}} t^{n-1}\eta_j\eta_k\xi_i, &(8.1)\cr
&\xi^{-1}\circ_{\hbox{h}}t^{n-1}\eta_i\eta_j\xi_j,\quad
\xi^{-1}\circ_{\hbox{h}} t^{n-1}\eta_k\eta_j\xi_k\xi_i, \cr
&n\xi^{-1}\circ_{\hbox{h}}t^{n-1}\eta_j\eta_k\xi_j\xi_k + \hbox{h}t^n,\quad
n\xi^{-1}\circ_{\hbox{h}} t^{n-1}\eta_j\eta_k\xi_i\xi_j\xi_k + 
\hbox{h}t^n\xi_i.\cr
}$$
Fix $\hbox{h} = 1$. Let $\mu \in (0, 1)$.
The action of the elements of $S$ on the spaces $V^{\mu}$
is well-defined. In each $V^{\mu}$ we defined a basis by (6.1).
We will denote it now by $V^{\mu} = <{v}_m^i(\mu), \hat{v}_m^i(\mu)>$.
Let $v(\mu)\in V^{\mu}$ be  vectors which have the same coordinates
with respect to this basis for all ${\mu}$.
Consider an odd nondegenerate superskew-symmetric form on each $V^{\mu}$:
$$\eqalignno{
&({v}_m^i(\mu), \hat{v}_l^i(\mu))_{\mu} = 
- (\hat{v}_l^i(\mu), {v}_m^i(\mu))_{\mu} = \delta_{m+l,0}
 \quad i = 1, 2, 3. &(8.2)\cr
&({v}_m^4(\mu), \hat{v}_l^4(\mu))_{\mu} =
- (\hat{v}_l^4(\mu), {v}_m^4(\mu))_{\mu} = -\delta_{m+l,0}.\cr
}$$
Let $V = <{v}_m^i, \hat{v}_m^i>$, where $i = 1, \ldots, 4$, $m\in\Z$,
be a superspace such that
$p({v}_m^i) = p(\hat{v}_m^4) = \bar{1}, 
p(\hat{v}_m^i) = p(v_m^4) = \bar{0}$.
A superskew-symmetric form on $V$ is defined by
$$\eqalignno{
&({v}_m^i, \hat{v}_l^i) = 
- (\hat{v}_l^i, {v}_m^i) = \delta_{m+l,0} \quad i = 1, 2, 3, &(8.3)\cr
&({v}_m^4, \hat{v}_l^4) =
- (\hat{v}_l^4, {v}_m^4) = -\delta_{m+l,0}.\cr
}$$
{\bf Theorem 5.}
$$\eqalignno{
&CK_6 = \lbrace X\in S \hbox{ }|\hbox{ } \lim_{\mu\rightarrow 0}
[(Xv(\mu), w(\mu))_{\mu} + (-1)^{p(X)p(v(\mu))}(v(\mu), Xw(\mu))_{\mu}] = 0,\cr
& \hbox{for all } v(\mu), w(\mu)\in V^{\mu}\rbrace. &(8.4)\cr
}$$
There is a representation of $CK_6$ in $V$  given by
(6.5), where $\mu = 0$, and this action preserves the form (8.3).
\hfil\break
{\bf Remark 2.} Correspondingly,
there is a representation of $CK_6$ in $V$  given by
(7.13), where $\mu = 0$, and this action preserves the odd nondegenerate
supersymmetric form on $V$:
$$({v}_m^i, \hat{v}_l^i) = 
 (\hat{v}_l^i, {v}_m^i) = \delta_{m+l,0} \quad i = 1, 2, 3, 4. \eqno(8.5)$$

\vfil\eject
\noindent
{\bf Acknowledgments}

This work is supported by the Anna-Greta and Holder Crafoords fond,
The Royal Swedish Academy of Sciences.
This work was started in  2002 at 
the Mathematical Sciences Research Institute, Berkeley,
during the program 
``Infinite-Dimensional Algebras and 
Mathematical Physics,'' and developed at the 
Institut des Hautes \'Etudes Scientifiques, Bures-sur-Yvette,
in  2003.
I wish to thank MSRI and IHES for the hospitality and support.
I am grateful to B. Feigin, V. Serganova and I. Shchepochkina for very 
useful discussions.
\vskip 0.1in
\noindent
{\bf References}
\font\red=cmbsy10
\def\~{\hbox{\red\char'0016}}

\item{[1]} M. Ademollo, L. Brink, A. D'Adda {\it et al}.,
``Dual strings with $U(1)$ colour symmetry'', 
Nucl. Phys. B {\bf 111}, 77-110 (1976).

\item{[2]} V. I. Arnold, 
{\it{Mathematical Methods of Classical Mechanics}}
(Springer-Verlag, New York, 1989).  

\item{[3]}
S.-J. Cheng and V. G.  Kac,
``A new $N = 6$ superconformal algebra'',
Commun. Math. Phys. {\bf 186},  219-231 (1997).

\item{[4]} B. Feigin and D. Leites,
``New Lie superalgebras of string theories'',
in {\it Group-Theore-
\hfil\break
tical Methods in
Physics}, edited by  M. Markov {\it et al.},
(Nauka, Moscow, 1983), Vol. 1, 269-273.
[English translation  Gordon and Breach, New York, 1984).

\item{[5]} B. Feigin,
{\it Private communication}.

\item{[6]} P. Grozman, D. Leites, and I. Shchepochkina,
``Lie superalgebras of string theories'',
Acta Math. Vietnam. {\bf 26}, no 1,  27-63 (2001).

\item{[7]}
V. G. Kac,
``Lie superalgebras'',
Adv. Math. {\bf 26}, 8-96 (1977).

\item{[8]}
V. G. Kac,
``Superconformal algebras and transitive group actions on quadrics'',
Commun. Math. Phys. {\bf 186}, 233-252 (1997).

\item{[9]}
V. G. Kac,
``Classification of infinite-dimensional simple linearly
compact Lie superalgebras'',
Adv. Math. {\bf 139}, 1-55 (1998).

\item{[10]}
V. G. Kac,
``Structure of some $\Z$-graded Lie superalgebras of vector fields'',
Transform. Groups {\bf 4}, 219-272 (1999).

\item{[11]}
V. G. Kac,
``Vertex algebras for beginners'',
University Lecture Series, Vol. 10 AMS, Providence, RI, 1996. 
Second edition 1998.

\item{[12]}
V. G. Kac and J. W. van de Leur,
``On classification of superconformal algebras'', in
{\it Strings-88}, edited by S. J. Gates {\it et al}.
(World Scientific, Singapore, 1989),  77-106.

\item{[13]}
B. Khesin, V. Lyubashenko, and C. Roger,
``Extensions and contractions of the Lie algebra of
q-pseudodifferential symbols on the circle'',
J. Funct. Anal. {\bf 143},  55-97 (1997).

\item{[14]}
O. S. Kravchenko and B. A. Khesin,
``Central extension of the algebra of 
pseudodifferential symbols'',
Funct. Anal. Appl. {\bf 25},  83-85 (1991).

\item{[15]}
V. Ovsienko and C. Roger,
``Deforming the Lie algebra of vector fields on $S^1$
inside the Poisson algebra on $\dot{T}^*S^1$'',
Commun. Math. Phys. {\bf 198}, 97-110 (1998).

\item{[16]}
V. Ovsienko and C. Roger,
``Deforming the Lie algebra of vector fields on $S^1$
inside the Lie algebra of pseudodifferential symbols on $S^1$'',
Am. Math. Soc. Trans. {\bf 194}, 211-226 (1999).

\item{[17]} I. Shchepochkina,
``The five exceptional simple Lie superalgebras of vector fields'',

hep-th/9702121.

\item{[18]} I. Shchepochkina,
``The five exceptional simple Lie superalgebras of vector fields'',
Funkt. Anal. i Prilozhen. {\bf 33},  59-72 (1999). 
[Funct. Anal. Appl. {\bf 33},  208-219 (1999)].

\item{[19]} I. Shchepochkina,
``The five exceptional simple Lie superalgebras of vector fields
and their fourteen regradings'',
Represent. Theory {\bf 3}, 373-415 (1999).

\item{[20]} E. Poletaeva, 
{\it A spinor-like representation of the contact
superconformal algebra  $K'(4)$}, 
J. Math. Phys. {\bf 42}, 526-540 (2001); hep-th/0011100
and references therein.
\bye